\documentclass{llncs}
\usepackage{graphicx}
\usepackage[usenames,dvipsnames,svgnames,table]{xcolor}
\usepackage{fancyvrb}
\usepackage{multicol}
\usepackage{upquote}
\usepackage{setspace}
\usepackage{hyperref}
\usepackage{listings}

\definecolor{linkblue}{RGB}{0,0,155}

\hypersetup{
   bookmarksnumbered,
   colorlinks={true},
   pdfstartview={FitH},
   citecolor={linkblue},
   linkcolor={linkblue},
   urlcolor={linkblue},
}

\newcommand{\code}[1]{\texttt{\small{}#1}}

\begin{document}

\title{Publishing without Publishers:\\a Decentralized Approach to Dissemination,\\Retrieval, and Archiving of Data}

\author{
Tobias Kuhn\inst{1,2}, Christine Chichester\inst{3}, Michael Krauthammer\inst{4}, and\\Michel Dumontier\inst{5}
}

\institute{
  Department of Humanities, Social and Political Sciences, ETH Zurich, Switzerland\\
\and
  Department of Computer Science, VU University Amsterdam, Netherlands\\
\and
  Swiss Institute of Bioinformatics, Geneva, Switzerland\\
\and
  Yale University School of Medicine, New Haven, CT, USA\\
\and
  Stanford Center for Biomedical Informatics Research,\\Stanford University, CA, USA
\smallskip\\
  \texttt{tokuhn@ethz.ch},
  \texttt{christine.chichester@isb-sib.ch},
  \texttt{michael.krauthammer@yale.edu},
  \texttt{michel.dumontier@stanford.edu}
}

\maketitle

\begin{abstract}
Making available and archiving scientific results is for the most part still considered the task of classical publishing companies, despite the fact that classical forms of publishing centered around printed narrative articles no longer seem well-suited in the digital age. In particular, there exist currently no efficient, reliable, and agreed-upon methods for publishing scientific datasets, which have become increasingly important for science.
Here we propose to design scientific data publishing as a Web-based bottom-up process, without top-down control of central authorities such as publishing companies. Based on a novel combination of existing concepts and technologies, we present a server network to decentrally store and archive data in the form of nanopublications, an RDF-based format to represent scientific data.
We show how this approach allows researchers to publish, retrieve, verify, and recombine datasets of nanopublications in a reliable and trustworthy manner, and we argue that this architecture could be used for the Semantic Web in general. Evaluation of the current small network shows that this system is efficient and reliable.
\end{abstract}

\section{Introduction}

Modern science increasingly depends on datasets, which however are left out in the classical way of publishing, i.e. through narrative (printed or online) articles in journals or conference proceedings. This means that the publications that describe scientific findings get disconnected from the data they are based on, which can seriously impair the verifiability and reproducibility of their results.
Addressing this issue raises a number of practical problems: How should one publish scientific datasets and how can one refer to them in the respective scientific publications? How can we be sure that the data will remain available in the future and how can we be sure that data we find on the Web have not been corrupted or tampered with? Moreover, how can we refer to specific entries or subsets from large datasets?

To address some of these problems, a number of scientific data repositories have appeared, such as Figshare and Dryad.\footnote{\url{http://figshare.com}, \url{http://datadryad.org}} Furthermore, Digital Object Identifiers (DOI) have been advocated to be used not only for articles but also for scientific data \cite{paskin2005digital}. While these services certainly improve the situation of scientific data, in particular when combined with Semantic Web techniques, they have nevertheless a number of drawbacks: They have centralized architectures, they give us no possibility to check whether the data have been (deliberately or accidentally) modified, and they do not support access or referencing on a more granular level than entire datasets (such as individual data entries).

Even if we put aside worst-case scenarios of organizations going bankrupt or becoming uninterested in sustaining their services, their websites have typically not a perfect uptime and might be down for a few minutes or even hours every once in a while. This is certainly acceptable for most use cases involving a human user accessing the data, but it can quickly become a problem in the case of automated access embedded in a larger service.
Furthermore, it is possible that somebody gains access to their database and silently modifies part of the data, or that the data get corrupted during the transfer from the server to the client.

Below we present an approach to tackle these problems, building upon existing Semantic Web technologies, in particular RDF and nanopublications, and adhering to accepted Web principles, such as decentralization and REST APIs.
Specifically, our research question is: Can we create a decentralized, reliable, trustworthy, and scalable system for publishing, retrieving, and archiving datasets in the form of sets of nanopublications based on existing Web standards and infrastructure?

\section{Background}
\label{sec:background}

Nanopublications \cite{groth2010isu} are a relatively recent proposal for improving the efficiency of finding, connecting, and curating scientific findings in a manner that takes attribution, quality levels, and provenance into account. While narrative articles would still have their place in the academic landscape, small formal data snippets in the form of nanopublications should take their central position in scholarly communication \cite{mons2011naturegen}. Most importantly, nanopublications can be automatically interpreted and aggregated and they allow for fine-grained citation metrics on the level of individual claims. On the technical level, nanopublications use the RDF language with named graphs \cite{carroll2005www} to represent assertions, as well as their provenance and metadata.
Conceptually, the approach boils down to the ideas of subdividing scientific results into atomic assertions, representing these assertions in RDF, attaching provenance information in RDF on the level of individual assertions, and treating each of these tiny entities as an individual publication.
Nanopublications have been applied to a number of domains, so far mostly from the life sciences including pharmacology \cite{williams2012open}, genomics \cite{patrinos2012humanmutation}, and proteomics \cite{chichester2014sw}.
An increasing number of datasets formatted as nanopublications are openly available, including neXtProt \cite{chichester2014querying} and DisGeNET \cite{queralt2014semanticweb}, and the nanopublication concept has been combined with and integrated into existing frameworks for data discovery and integration, such as CKAN \cite{mccusker2013next}.

Research Objects are a related proposal to establish ``self-contained units of knowledge'' \cite{belhajjame2012sepublica}, and they constitute in a sense the antipode approach to nanopublications. We could call them ``megapublications,'' as they contain much more than a typical narrative publication, namely resources like input and output data, workflow definitions, log files, and presentation slides.
We demonstrate in this paper, however, that bundling all resources of scientific studies in large packages is not a necessity to ensure reproducibility and trust, but we can achieve these properties also with strong identifiers and a decentralized server network.

SPARQL endpoints, i.e. query APIs to RDF triple stores, are a widely used technique for making linked data available on the Web in a flexible manner. While off-the-shelf triple stores can nowadays handle billions of triples or more, they require a significant amount of resources in the form of memory and processor time to do so, at least if the full expressive power of the SPARQL language is supported.
A recent study found that more than half of the publicly accessible SPARQL endpoints are available less than 95\% of the time \cite{builaranda2013iswc}, posing a major problem to services depending on them, in particular to those that depend on several endpoints at the same time.
To solve these problems, alternative approaches and platforms --- such as Linked Data Fragments \cite{verborgh2014ldow}, the Linked Data Platform \cite{speicher2015ldp}, and CumulusRDF \cite{ladwig2011ssws} --- have been proposed, providing less powerful query interfaces and thereby shifting the workload from the server to the client.

Fully reliable services, however, can only be achieved with distributed architectures, which have been proposed by a number of existing approaches related to data publishing.
For example, distributed file systems that are based on cryptographic methods have been designed for data that are public \cite{fu2002acm} or private \cite{clarke2001freenet}.
In contrast to the design principles of the Semantic Web, these approaches implement their own internet protocols and follow the hierarchical organization of file systems.
Other approaches build upon the existing BitTorrent protocol and apply it to data publishing \cite{markman2014dlib,cohen2014xsede}, and there is interesting work on repurposing the proof-of-work tasks of Bitcoin for data preservation \cite{miller2014sp}.
There exist furthermore a number of approaches to applying peer-to-peer networks for RDF data \cite{filali2011lsdkcs}, but they do not allow for the kind of permanent and provenance-aware publishing that we propose below.
Moreover, only for the centralized and closed-world setting of database systems, approaches exist that allow for robust and granular references to subsets of dynamic datasets \cite{proell2014data}.

Our approach is based on previous work, in which we proposed \emph{trusty URIs} to make nanopublications and their entire reference trees verifiable and immutable by the use of cryptographic hash values \cite{kuhn2014eswc,kuhn2015tkde}. This is an example of such a trusty URI:
\begin{lstlisting}[basicstyle=\small\ttfamily]
http://example.org/r1.RA5AbXdpz5DcaYXCh9l3eI9ruBosiL5XDU3rxBbBaUO70
\end{lstlisting}
The last 45 characters of this URI (i.e. everything after ``\code{.}'') is what we call the \emph{artifact code}. It contains a hash value that is calculated on the RDF content it represents, such as the RDF graphs of a nanopublication. Because this hash is part of the URI, any link to such an artifact comes with the possibility to verify its content, including other trusty URI links it might contain. In this way, the range of verifiability extends to the entire reference tree.

Furthermore, we argued in previous work that the assertion of a nanopublication need not be fully formalized, but we can allow for informal or underspecified assertions \cite{kuhn2013eswc}.
We also sketched how ``science bots'' could autonomously produce and publish nanopublications, and how algorithms could thereby be tightly linked to their generated data \cite{kuhn2015savesd}, which requires the existence of a reliable and trustworthy publishing system, such as the one we present here.

\section{Approach}

Our approach builds upon the existing concept of nanopublications and our previously introduced method of trusty URIs.
It is a proposal of a reliable implementation of accepted Semantic Web principles, in particular of what has become known as the \emph{follow-your-nose} principle: Looking up a URI should return relevant data and links to other URIs, which allows one (i.e. humans as well as machines) to discover things by navigating through this data space \cite{berners2006linked}. We argue that approaches following this principle can only be reliable and efficient if we have some sort of guarantee that the resolution and processing of any single identifier will succeed in one way or another and only takes up a small amount of time and resources.
This requires (1) that RDF representations are made available on several distributed servers, so the chance that they all happen to be inaccessible at the same time is negligible, and that (2) these representations are reasonably small, so that downloading them is a matter of fractions of a second, and so that one has to process only a reasonable amount of data to decide which links to follow.
We address the first requirement by proposing a distributed server network and the second one by building upon the concept of nanopublications.
Below we explain the general architecture, the functioning and the interaction of the nanopublication servers, and the concept of nanopublication indexes.

\subsection{Architecture}
\label{sec:architecture}

\begin{figure}[t]
\begin{center}
\includegraphics[width=\textwidth]{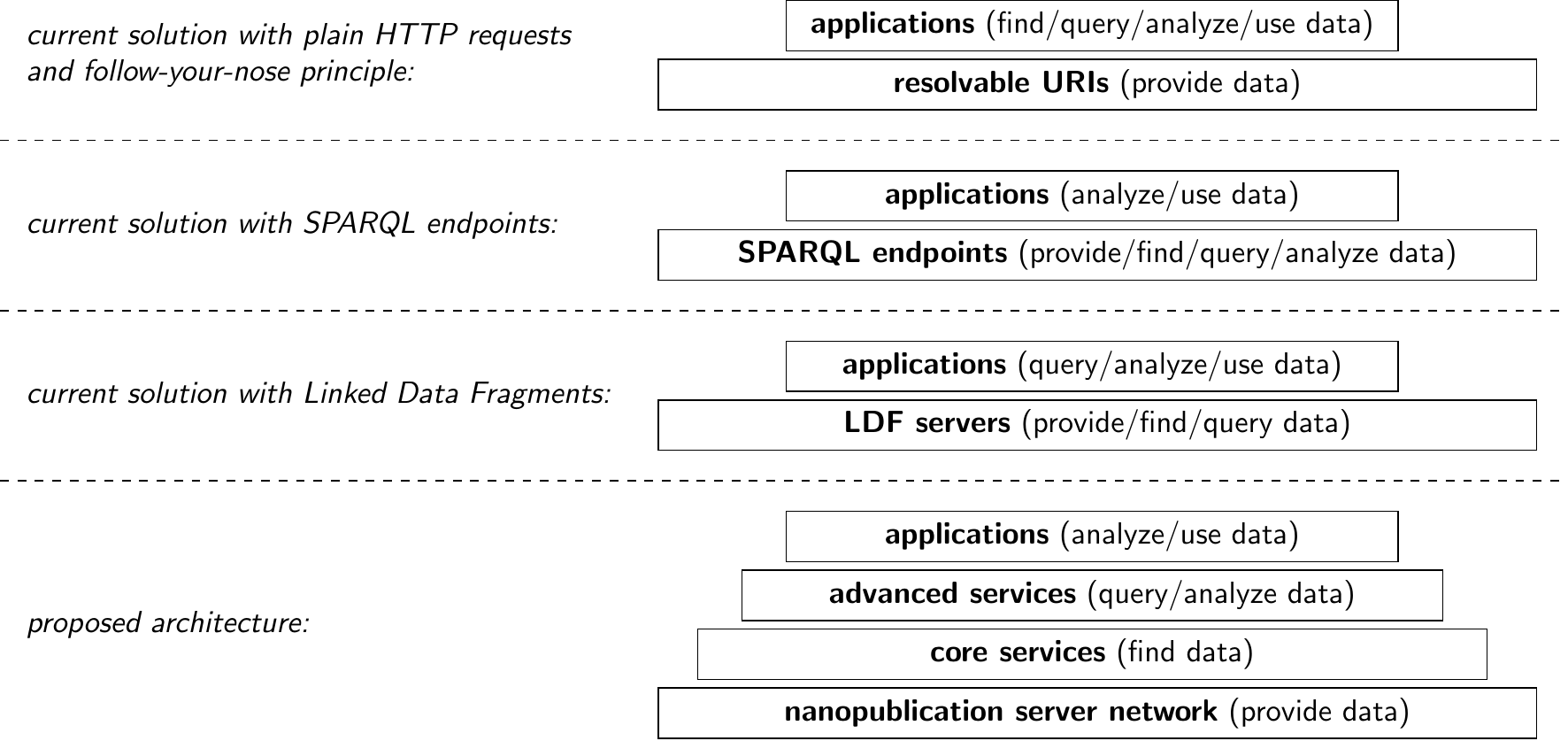}
\caption{Illustration of current architectures of Semantic Web applications and our proposed approach}
\label{fig:swarch}
\end{center}
\end{figure}

There are currently at least three possible architectures for Semantic Web applications (and mixtures thereof), as shown in a simplified manner in Figure \ref{fig:swarch}.
The first option is the use of plain HTTP GET requests. Applying the follow-your-nose principle, resolvable URIs provide the data based on which the application performs the tasks of finding relevant resources, running queries, analyzing and aggregating the results, and using them for the purpose of the application.
If SPARQL endpoints are used, as a second option, most of the workload is shifted from the application to the server via the expressive power of the SPARQL query language.
A more reasonable approach, in our view, is the third option of Linked Data Fragments, where servers provide only limited query features and where the tasks are distributed between servers and applications in more balanced fashion.
However, all these current solutions are based on two-layer architectures, and have moreover no inherent replication mechanisms. A single point of failure can cause applications to be unable to complete their tasks: A single URI that does not resolve or a single server that does not respond can break the entire process.

We argue here that we need distributed and decentralized services to allow for robust and reliable applications that consume linked data. At the same time, the most low-level task of providing linked data is essential for all other tasks at higher levels, and therefore needs to be the most stable and robust one. We argue that this can be best achieved if we free this lowest layer from all tasks except the provision and archiving of data entries (nanopublications in our case) and decouple it from the tasks of providing services for finding, querying, or analyzing the data. This makes us advocate a multi-layer architecture, a possible realization of which is shown at the bottom of Figure \ref{fig:swarch}.

Below we present a concrete proposal of such a low-level data provision infrastructure in the form of a nanopublication server network. Based on such an infrastructure, one can then build different kinds of services operating on a subset of the nanopublications they find in the underlying network. ``Core services'' could involve things like resolving backwards references (i.e. ``which nanopublications refer to the given one?'') and the retrieval of the nanopublications published by a given person or containing a particular URI. Based on such core services for finding nanopublications, one could then provide ``advanced services'' that allow us to run queries on subsets of the data and ask for aggregated output.
(These higher layers could of course make use of existing techniques such as SPARQL endpoints and Linked Data Fragments.)
While the lowest layer would necessarily be accessible to everybody, some of the services on the higher level could be private or limited to a small (possibly paying) user group.
We have in particular scientific data in mind, but we think that an architecture of this kind could also be used for Semantic Web content in general.

\subsection{Nanopublication Servers}

As a concrete proposal of a low-level data provision layer, as explained above, we present here a decentralized nanopublication server network with a REST API to provide and propagate nanopublications identified by trusty URIs.\footnote{\url{https://github.com/tkuhn/nanopub-server}%
} The nanopublication servers of such a network connect to each other to retrieve and replicate their nanopublications, and they allow users to upload new nanopublications, which are then automatically distributed through the network.


Basing the content of this network on nanopublications with trusty URIs has a number of positive consequences for its design: The first benefit is that the fact that nanopublications are all similar in size and always small makes it easy to estimate how much time is needed to process an entity (such as validating its hash) and how much space to store it (e.g. as a serialized RDF string in a database). Moreover it ensures that these processing times remain mostly in the fraction-of-a-second range, guaranteeing quick responses, and that these entities are never too large to be analyzed in memory.
The second benefit is that servers do not have to deal with identifier management, as the nanopublications already come with trusty URIs, which are guaranteed to be unique and universal. The third and possibly most important benefit is that nanopublications with trusty URIs are immutable and verifiable. This means that servers only have to deal with \emph{adding} new entries but not with \emph{updating} or \emph{correcting} any of them, which eliminates the hard problems of concurrency control and data integrity in distributed systems. Together, these aspects significantly simplify the design of such a network and its synchronization protocol, and make it reliable and efficient even with limited resources.

Specifically, a nanopublication server of the current network has the following components:
\begin{itemize}
\item A \textbf{key-value store} of its nanopublications (with the trusty URI as the key)
\item A \textbf{journal} consisting of a journal identifier and a list of the identifiers of all loaded nanopublications, subdivided into pages of a fixed size. 
\item Optionally, a \textbf{cache of gzipped packages} containing all nanopublications for a given journal page (but they can also be generated on the fly)
\item A \textbf{list of known peers}, i.e. the URLs of other nanopublication servers
\item \textbf{Information about each known peer}, including the journal identifier and the total number of nanopublications at the time it was last visited
\end{itemize}
Based on these components, the servers respond to the following request (in the form of HTTP GET):
\begin{itemize}
\item Each server needs to return general \textbf{server information}, including the journal identifier and the number of stored nanopublications
\item Given an artifact code (i.e. the final part of a trusty URI) of a known nanopublication, the server returns the given \textbf{nanopublication} in a format like TriG, TriX, or N-Quads (depending on content negotiation).
\item A \textbf{journal page} can be requested by page number as a list of trusty URIs.
\item For every journal page (except for incomplete last pages), a gzipped \textbf{package} can be requested containing the respective nanopublications.
\item The \textbf{list of known peers} can be requested as a list of URLs.
\end{itemize}
In addition, a server can optionally support the following two actions (in the form of HTTP POST requests):
\begin{itemize}
\item A server may accept requests to \textbf{add a given individual nanopublication} to its database.
\item A server may also accept requests to \textbf{add the URL of a new nanopublication server} to its peer list.
\end{itemize}
Server administrators have the additional possibility to load nanopublications from the local file system.
Together, these server components and their possible interactions allow for efficient decentralized distribution of published nanopublications.

The current system can be seen as an unstructured peer-to-peer network, where each node can freely decide which other nodes to connect to and which nanopublications to replicate. As the network is still very small, the present five nodes connect to all other nodes and replicate all nanopublications they can find.
The current implementation is furthermore designed to be run on normal Web servers alongside with other applications, with economic use of the server's resources in terms of memory and processing time. In order to avoid overload of the server or the network connection, we restrict outgoing connections to other servers to one at a time.
The current system and its protocol are not set in stone but, if successful, will have to evolve in the future --- in particular with respect to network topology and partial replication --- to accommodate a network of possibly thousands of servers and billions of nanopublications.

\subsection{Nanopublication Indexes}

To make the infrastructure described above practically useful, we have to introduce the concept of indexes.
One of the core ideas behind nanopublications is that each of them is a tiny atomic piece of data. This implies that analyses will mostly involve more than just one nanopublication and typically a large number of them. Similarly, most processes will generate more than just one nanopublication, possibly thousands or even millions of them. Therefore, we need to be able to group nanopublications and to identify and use large collections of them.

Given the versatility of the nanopublication standard, it seems straightforward to represent such collections as nanopublications themselves. However, if we let such ``collection nanopublications'' contain other nanopublications, then the former would become very large for large collections and would quickly lose their property of being \emph{nano}.
We can solve part of that problem by applying a principle that we can call \emph{reference instead of containment}: nanopublications cannot contain but only refer to other nanopublications, and trusty URIs allow us to make these reference links almost as strong as containment links.
To emphasize this principle, we call them \emph{indexes} and not collections.

\begin{figure}[t]
\begin{center}
\includegraphics[width=0.85\textwidth]{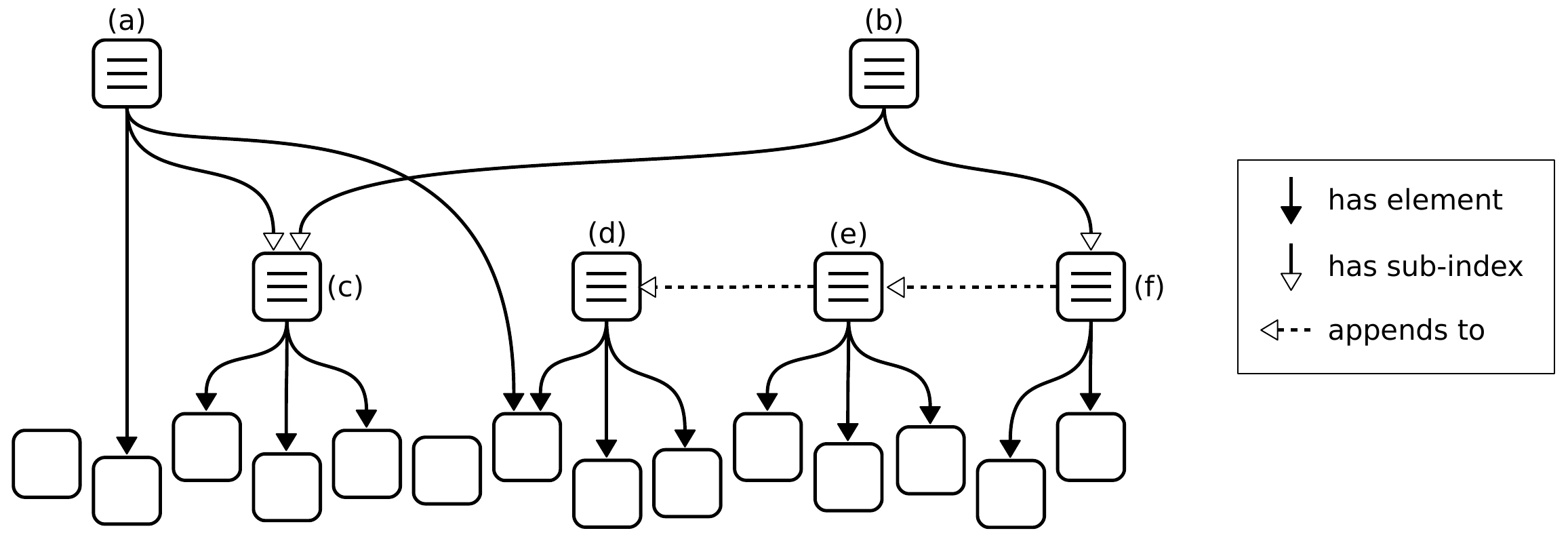}
\caption{Schematic example of nanopublication indexes
}
\label{fig:indexes}
\end{center}
\end{figure}

However, even by only containing references and not the complete nanopublications, these indexes can still become quite large. To ensure that all such index nanopublications remain \emph{nano} in size%
, we need to put some limit on the number of references, and to support sets of arbitrary size, we can allow indexes to be appended by other indexes.
We set 1000 nanopublication references as the upper limit any single index can directly contain. This limit is admittedly arbitrary, but it seems to be a reasonable compromise between ensuring that nanopublications remain small on the one hand and limiting the number of nanopublications needed to define large indexes on the other.
A set of 100,000 nanopublications, for example, can therefore be defined by a sequence of 100 indexes, where the first one stands for the first 1000 nanopublications, the second one appends to the first and adds another 1000 nanopublications (thereby representing 2000 of them), and so on up to the last index, which appends to the second to last and thereby stands for the entire set. In addition, to allow datasets to be organized in hierarchies, we define that the references of an index can also point to sub-indexes.
In this way we end up with three types of relations: an index can \emph{append to} another index, it can contain other indexes as \emph{sub-indexes}, and it can contain nanopublications as \emph{elements}.
These relations defining the structure of nanopublication indexes are shown schematically in Figure~\ref{fig:indexes}.
Index (a) in the shown example contains five nanopublications, three of them via sub-index (c). The latter is also part of index (b), which additionally contains eight nanopublications via sub-index (f). Two of these eight nanopublications belong directly to (f), whereas the remaining six come from appending to index (e). Index (e) in turn gets half of its nanopublications by appending to index (d).
We see that some nanopublications may not be referenced by any index at all, while others may belong to several indexes at the same time.

Below we show how this general concept of indexes can be used to define sets of new or existing nanopublications, and how such index nanopublications can be published and their nanopublications retrieved.

\subsection{Trusty Publishing}

Let us consider two simple exemplary scenarios to illustrate and motivate the general concepts, using the \code{np} command from the \code{nanopub-java} library\footnote{\url{https://github.com/Nanopublication/nanopub-java}}.
Given, for example, a file \code{nanopubs.trig} with three nanopublications, we have to assign them trusty URIs before they can be published:
\begin{lstlisting}
{c}$ np mktrusty -v nanopubs.trig
{o}Nanopub URI: http://example.org/np1#RAQoZlp22LHIvtYqHCosPbUtX8yeGs1Y5AfqcjMneLQ2I
{o}Nanopub URI: http://example.org/np2#RAT5swlSLyMbuD03KzJsYHVV2oM1wRhluRxMrvpkZCDUQ
{o}Nanopub URI: http://example.org/np3#RAkvUpysi9Ql3itlc6-iIJMG7YSt3-PI8dAJXcmafU71s{end}
\end{lstlisting}
This gives us the file \code{trusty.nanopubs.trig}, which contains transformed versions of the three nanopublications, now having trusty URIs as identifiers. We can now publish these nanopublications to the network:
\begin{lstlisting}
{c}$ np publish trusty.nanopubs.trig
{o}3 nanopubs published at http://np.inn.ac/{end}
\end{lstlisting}
We can check the publication status of the given nanopublications:
\begin{lstlisting}
{c}$ np status -a http://example.org/np1#RAQoZlp22LHIvtYqHCosPbUtX8yeGs1Y5AfqcjMneLQ2I
{o}URL: http://np.inn.ac/RAQoZlp22LHIvtYqHCosPbUtX8yeGs1Y5AfqcjMneLQ2I
{o}Found on 1 nanopub server.{end}
\end{lstlisting}
This is what we see immediately after publication, but only a few minutes later the given nanopublication is found on several servers:
\begin{lstlisting}
{c}$ np status -a http://example.org/np1#RAQoZlp22LHIvtYqHCosPbUtX8yeGs1Y5AfqcjMneLQ2I
{o}URL: http://np.inn.ac/RAQoZlp22LHIvtYqHCosPbUtX8yeGs1Y5AfqcjMneLQ2I
{o}URL: http://ristretto.med.yale.edu:8080/nanopub-server/RAQoZlp22LHIvtYqHCosPbUtX8yeGs{k}...
{o}URL: http://nanopub-server.ops.labs.vu.nl/RAQoZlp22LHIvtYqHCosPbUtX8yeGs1Y5AfqcjMneLQ2I
{o}URL: http://nanopubs.stanford.edu/nanopub-server/RAQoZlp22LHIvtYqHCosPbUtX8yeGs1Y5Afq{k}...
{o}URL: http://nanopubs.semanticscience.org/RAQoZlp22LHIvtYqHCosPbUtX8yeGs1Y5AfqcjMneLQ2I
{o}Found on 5 nanopub servers.{end}
\end{lstlisting}
Next, we can make an index pointing to these three nanopublications:
\begin{lstlisting}
{c}$ np mkindex -o index.nanopubs.trig trusty.nanopubs.trig 
{o}Index URI: http://np.inn.ac/RAXsXUhY8iDbfDdY6sm64hRFPr7eAwYXRlSsqQAz1LE14{end}
\end{lstlisting}
This creates a local file \code{index.nanopubs.trig} containing the index, identified by the URI shown above. As this index is itself a nanopublication, we can publish it in the same way as described above, and then everybody can conveniently and reliably retrieve the given set of nanopublications:
\begin{lstlisting}
{c}$ np get -c http://np.inn.ac/RAXsXUhY8iDbfDdY6sm64hRFPr7eAwYXRlSsqQAz1LE14{end}
\end{lstlisting}
This command downloads the content of the given index, i.e. the three nanopublications we just created and published.

As another exemplary scenario, let us imagine a researcher in the biomedical domain who is interested in the protein CDKN2A and who has derived some conclusion based on the data found in existing nanopublications. Specifically, let us suppose this researcher analyzed five nanopublications from different sources, specified by the following artifact codes (they can be viewed online by appending the artifact code to the URL \nolinkurl{http://np.inn.ac/}):
\begin{lstlisting}
{o}RAEoxLTy4pEJYbZwA9FuBJ6ogSquJobFitoFMbUmkBJh0
{o}RAoMW0xMemwKEjCNWLFt8CgRmg_TGjfVSsh15hGfEmcz4
{o}RA3BH_GncwEK_UXFGTvHcMVZ1hW775eupAccDdho5Tiow
{o}RA3HvJ69nO0mD5d4m4u-Oc4bpXlxIWYN6L3wvB9jntTXk
{o}RASx-fnzWJzluqRDe6GVMWFEyWLok8S6nTNkyElwapwno{end}
\end{lstlisting}
%
These nanopublications can be downloaded from the network with the \code{np get} command and stored in a file, which we name here \code{cdkn2a-nanopubs.trig}.
In order to be able to refer to such a collection of nanopublications with a single identifier, a new index is needed that refers to just these five nanopublications. This time we give the index a title (which is optional):
\begin{lstlisting}
{c}$ np mkindex -t "Data about CDKN2A from BEL2nanopub & neXtProt" \
{c}  -o index.cdkn2a-nanopubs.trig cdkn2a-nanopubs.trig
{o}Index URI: http://np.inn.ac/RA6jrrPL2NxxFWlo6HFWas1ufp0OdZzS_XKwQDXpJg3CY{end}
\end{lstlisting}
The generated index is stored in the file \code{index.cdkn2a-nanopubs.trig}, and our exemplary researcher can now publish this index to let others know about it:
\begin{lstlisting}
{c}$ np publish index.cdkn2a-nanopubs.trig
{o}1 nanopub published at http://np.inn.ac/{end}
\end{lstlisting}
There is no need to publish the five nanopublications this index is referring to, because they are already public (this is how we got them in the first place). The index URI can be used to refer to this new collection of existing nanopublications in an unambiguous and reliable manner, for example as a reference in a paper, as we do it for the datasets of this article \cite{nanopubindex2015aidagenerif,nanopubindex2015openbel1,nanopubindex2015openbel2,nanopubindex2015disgenet,nanopubindex2015nextprot}.



\section{Evaluation}
\label{sec:eval}

\begin{figure*}[t]
\begin{center}
\includegraphics[width=0.95\textwidth]{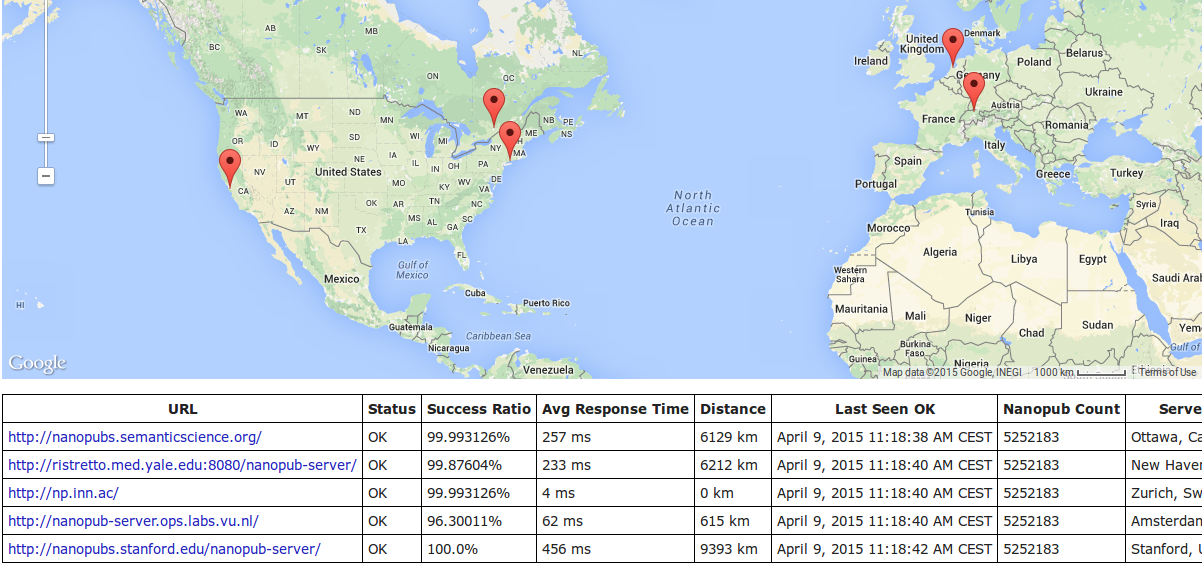}
\caption{This screenshot of the nanopublication monitor interface (\url{http://npmonitor.inn.ac}) showing the current server network. 
}
\label{fig:map}
\end{center}
\end{figure*}

To evaluate our approach, we want to find out whether a small server network run on normal Web servers, without dedicated infrastructure, is able to handle the amount of nanopublications we can expect to become publicly available in the next few years.
%
At the time the evaluation was performed, the server network consisted of three servers in Zurich, New Haven, and Ottawa. Two new servers in Amsterdam and Stanford have joined the network since. The current network of five servers is shown in Figure~\ref{fig:map}, which is a screenshot of a nanopublication monitor that we have implemented.
Such monitors regularly check the nanopublication server network, register changes (currently once per minute), and test the response times and the correct operation of the servers by requesting a random nanopublication and verifying the returned data. 

\subsection{Evaluation Design}

\begin{table*}[t]
\begin{center}
{\footnotesize
\begin{tabular*}{\textwidth}{@{\extracolsep{\stretch{1}}}*{6}{l|rr|rr|l}@{}}
 & \multicolumn{2}{c|}{\# nanopubs} & \multicolumn{2}{c|}{\# triples} & initial location \\
dataset & \emph{index} & \emph{content} & \emph{index} & \emph{content} & for evaluation \\
\hline
GeneRIF/AIDA \cite{nanopubindex2015aidagenerif} & 157 & 156,026 & 157,909 & 2,340,390 & New Haven \\
OpenBEL 1.0 \cite{nanopubindex2015openbel1} & 53 & 50,707 & 51,448 & 1,502,574 & New Haven \\
OpenBEL 20131211 \cite{nanopubindex2015openbel2} & 76 & 74,173 & 75,236 & 2,186,874 & New Haven \\
DisGeNET v2.1.0.0 \cite{nanopubindex2015disgenet} & 941 & 940,034 & 951,325 & 31,961,156 & Zurich \\
neXtProt \cite{nanopubindex2015nextprot} & 4,026 & 4,025,981 & 4,078,318 & 156,263,513 & Ottawa \\
\hline
total & 5,253 & 5,246,921 & 5,314,236 & 194,254,507 & \\
\end{tabular*}
}
\caption{Existing datasets in the nanopublication format that were used for the first part of the evaluation.}
\label{tab:datasets}
\end{center}
\end{table*}

Table \ref{tab:datasets} shows the existing datasets that we use for the first part of the evaluation.
This includes all datasets we are aware of that use trusty URIs, with a total of more than 5 million nanopublications and close to 200 million RDF triples, including nanopublication indexes that we generated for each dataset. The total size of these datasets when stored as uncompressed TriG files amounts to 15.6 GB. Each of the datasets is assigned to one of the three servers, where it is loaded from the local file systems. The first nanopublications start spreading to the other servers, while others are still being loaded from the file system.
We therefore test the reliability and capacity of the network under constant streams of new nanopublications coming from different servers, and we use two nanopublication monitors (in Zurich and Ottawa) to evaluate the responsiveness of the network.

In the second part of the evaluation we expose a server to heavy load from clients to test its retrieval capacity. For this we use a service called Load Impact\footnote{\url{https://loadimpact.com}} to let up to 100 clients access a nanopublication server in parallel. We test the server in Zurich over a time of five minutes under the load from a linearly increasing number of clients (from 0 to 100) located in Dublin. These clients are programmed to request a randomly chosen journal page, then to go though the entries of that page one by one, requesting the respective nanopublication with a probability of 10\%, and starting over again with a different page.
As a comparison, we run a second session, for which we load the same data into a Virtuoso SPARQL endpoint on the same server in Zurich (with 16 GB of memory given to Virtuoso and two 2.40 GHz Intel Xeon processors). Then, we perform exactly the same stress test on the SPARQL endpoint, requesting the nanopublications in the form of SPARQL queries instead of requests to the nanopublication server interface.
This comparison is admittedly not a fair one, as SPARQL endpoints are much more powerful and are not tailor-made for the retrieval of nanopublications, but they provide nevertheless a valuable and well-established reference point to evaluate the performance of our system.

\subsection{Evaluation Results}

\definecolor{chartred}{RGB}{192,47,47}
\definecolor{chartgreen}{RGB}{47,192,47}
\definecolor{chartblue}{RGB}{47,47,192}

\begin{figure*}[t]
\begin{center}
\resizebox{\textwidth}{!}{
\begin{minipage}{18mm}\small
\begin{flushleft}\sffamily
\textbf{\color{chartgreen}{Zurich:}}\\
~\\
~\\
~\\
\textbf{\color{chartred}{New Haven:}}\\
~\\
~\\
~\\
\textbf{\color{chartblue}{Ottawa:}}\\
~\\
~\\
\vspace{8mm}
\end{flushleft}
\end{minipage}%
\begin{minipage}{23mm}\small
\begin{flushright}\sffamily
loaded locally\\
from New Haven\\
from Ottawa\\
~\\
loaded locally\\
from Zurich\\
from Ottawa\\
~\\
loaded locally\\
from Zurich\\
from New Haven\\
\vspace{8mm}
\end{flushright}
\end{minipage}~%
\begin{minipage}{120mm}
\includegraphics[height=42mm,width=120mm]{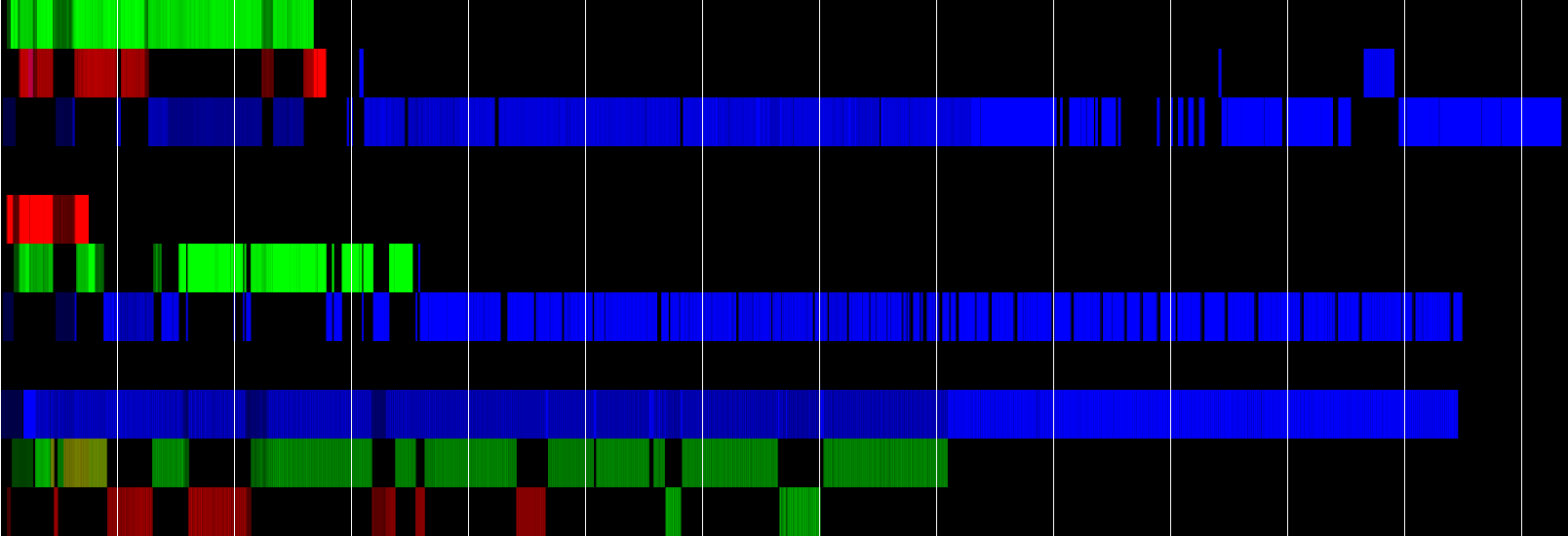}\\
\sffamily\small
0\phantom{0} \hfill 1\phantom{0} \hfill 2\phantom{0} \hfill 3\phantom{0} \hfill 4\phantom{0} \hfill 5\phantom{0} \hfill 6\phantom{0} \hfill 7\phantom{0} \hfill 8\phantom{0} \hfill 9\phantom{0} \hfill 10 \hfill 11 \hfill 12 \hfill 13 ~\\
\centering{time from start of evaluation in hours}
\end{minipage}
}
\caption{The flow of nanopublications during the time of the evaluation.
The colors indicate the \emph{original} location of the respective nanopublications, and the brightness stands for the rate at which they are loaded (bright meaning high rate).}
\label{fig:dataflow}
\end{center}
\end{figure*}

The first part of the evaluation lasted 13 hours and 21 minutes, at which point all nanopublications were replicated on all three servers, and therefore the nanopublication traffic came to an end.
Figure \ref{fig:dataflow} shows the type and intensity of the data flow (i.e. the transfer of nanopublications) between the three servers over the time of the evaluation.
%
%
The network was able to handle an average of about 400,000 new nanopublications per hour, which corresponds to more than 100 new nanopublications per second. This includes the time needed for loading each nanopublication once from the local file system (at the first server), transferring it through the network two times (to the other two servers), and for verifying it three times (once when loaded and twice when received by the other two servers).
\begin{figure*}[t]
\begin{center}
\begin{minipage}{0.02\textwidth}
\scalebox{0.6}{\rotatebox{90}{\sffamily ~ ~ ~ \emph{response times in milliseconds:}}}
\end{minipage}%
\begin{minipage}{0.98\textwidth}
\includegraphics[trim=0mm 10mm 0mm 0mm, clip=true, width=\textwidth]{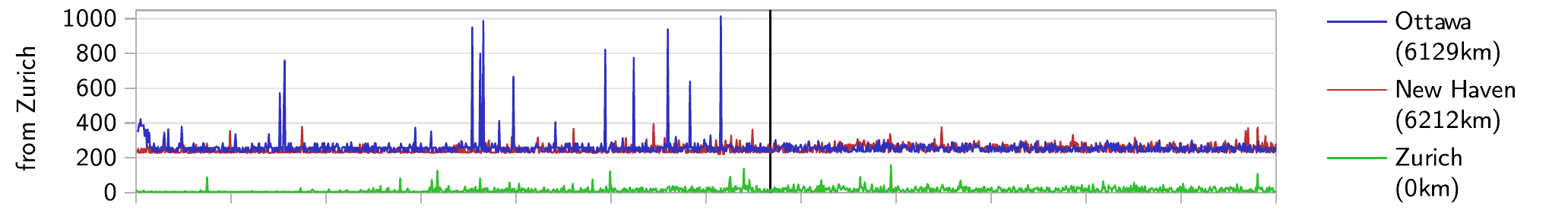}\\
\includegraphics[width=\textwidth]{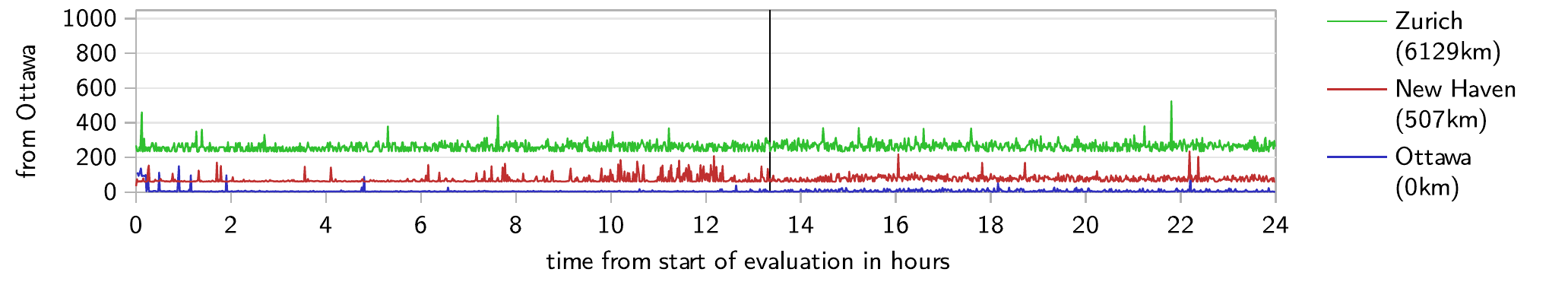}
\end{minipage}
\caption{Server response times as recorded during and after the first evaluation, which ended at 13 hours and 21 minutes, as indicated by the black vertical line.}
\label{fig:responsetimes}
\end{center}
\end{figure*}
Figure \ref{fig:responsetimes} shows the response times of the three servers as measured by the two nanopublication monitors in Zurich (top) and Ottawa (bottom) from the start of the evaluation until 24 hours later, therefore covering the entire evaluation plus an additional 10 hours and 39 minutes after its end.
We see that the observed latency is mostly due to the geographical distance between the servers and the monitors. The response time was always less than 0.25 seconds when the server was on the same continent as the measuring monitor. In 99.86\% of all cases (including those across continents) the response time was below 0.5 seconds, and it was always below 1.1 seconds. Not a single one of the 8636 individual HTTP requests timed out, led to an error, or received a nanopublication that could not be successfully verified.
We see that the load put onto the network did not have much of an impact on the response times. Except for a handful of spikes, one barely notices the difference between the heavy-load and zero-load situations.

\begin{figure*}[t]
\begin{center}
\includegraphics[trim=0mm 6mm 0mm 0mm, clip=true, width=0.9\textwidth]{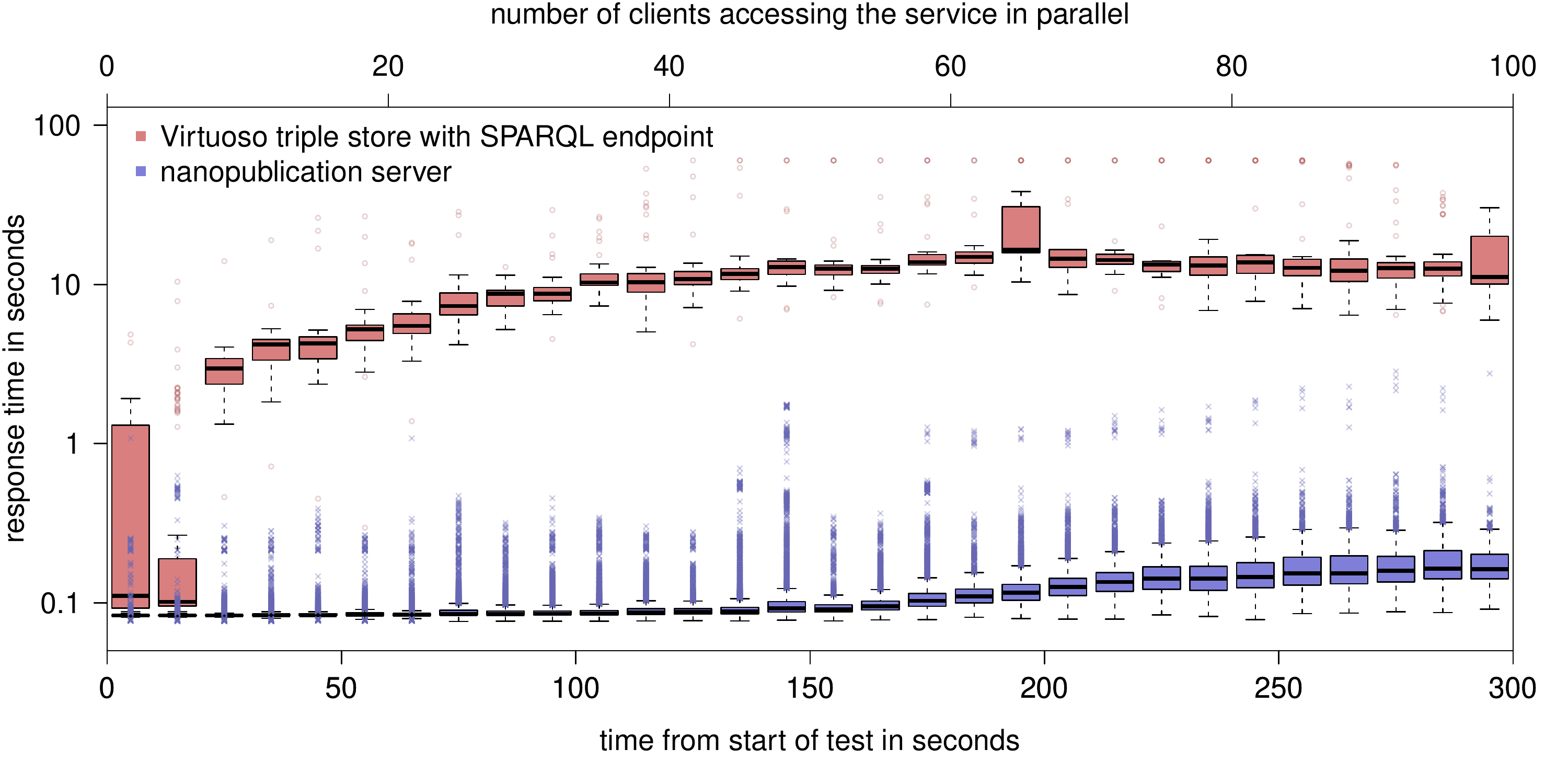}
\caption{Results of the evaluation of the retrieval capacity of a nanopublication server as compared to a general SPARQL endpoint (note the logarithmic $y$-axis)}
\label{fig:loadimpact}
\end{center}
\end{figure*}

Figure \ref{fig:loadimpact} shows the result of the second part of the evaluation. The nanopublication server was able to handle 113,178 requests in total (i.e. an average of 377 requests per second) with an average response time of 0.12 seconds. In contrast, the SPARQL endpoint answering the same kind of requests needed 100 times longer to process them (13 seconds on average), consequently handled about 100 times fewer requests (1267), and started to hit the timeout of 60 seconds for some requests when more than 40 client accessed it in parallel.
In the case of the nanopublication server, the majority of the requests were answered within less than 0.1 seconds for up to around 50 parallel clients, and this value remained below 0.17 seconds all the way up to 100 clients. As the round-trip network latency alone between Ireland and Zurich amounts to around 0.03 to 0.04 seconds, further improvements can be achieved for a denser network due to the reduced distance to the nearest server.

The first part of the evaluation shows that the overall replication capacity of the current server network is around 9.4 million new nanopublications per day or 3.4 billion per year.
The results of the second part show that the load on a server when measured as response times is barely noticeable for up to 50 parallel clients, and therefore the network can easily handle $50 \cdot x$ parallel client connections or more, where $x$ is the number of servers in the network (currently $x = 5$).
The second part thereby also shows that the restriction of avoiding parallel outgoing connections for the replication between servers is actually a very conservative measure that could be relaxed, if needed, to allow for a higher replication capacity.

\section{Discussion and Conclusion}

We have presented here a low-level infrastructure for data sharing, which is just one piece of a bigger ecosystem to be established. The implementation of components that rely on this low-level data sharing infrastructure is ongoing and future work. This includes the development of ``core services'' (see Section \ref{sec:architecture}) on top of the server network to allow people to find nanopublications and ``advanced services'' to query and analyze the content of nanopublications.
In addition, we need to establish standards and best practices of how to use existing ontologies (and to define new ones where necessary) to describe properties and relations of nanopublications, such as referring to earlier versions, marking nanopublications as retracted, and reviewing of nanopublications.

Apart from that, we also have to scale up the current small network. As our protocol only allows for simple key-based lookup, the time complexity for all types of requests is sublinear and therefore scales up well. The main limiting factor is disk space, which is relatively cheap and easy to add. Still, the servers will have to specialize, i.e. replicate only a part of all nanopublications, in order to handle really large amounts of data, which can be done in a number of ways:
Servers can restrict themselves to nanopublications from a certain internet domain, or to particular types of nanopublications, e.g. to specific topics or authors, and communicate this to the network;
inspired by the Bitcoin system, certain servers could only accept nanopublications whose hash starts with a given number of zero bits, which makes it costly to publish; and
some servers could be specialized to new nanopublications, providing fast access but only for a restricted time, while others could take care of archiving old nanopublications, possibly on tape and with considerable delays between request and delivery.
Lastly, there could also emerge interesting synergies with novel approaches to internet networking, such as Content-Centric Networking \cite{jacobson2012acm}, with which --- consistent with our proposal --- requests are based on content rather than hosts.


We argue that data publishing and archiving can and should be done in a decentralized manner.
We believe that the presented server network can serve as a solid basis for semantic publishing, and possibly also for the Semantic Web in general. It could contribute to improve the availability and reproducibility of scientific results and put a reliable and trustworthy layer underneath the Semantic Web.

\bibliographystyle{abbrv}
\bibliography{trustypublishing}

\end{document}